\documentclass[12pt]{iopart}
\usepackage{epsfig}
\usepackage{graphicx}
\usepackage{dcolumn}
\usepackage{bm}
\usepackage{iopams}
\usepackage{hyperref}
\usepackage{gensymb}

\newcommand{\zh}{\bm}

\newcommand{\trace}{\mathop{\rm Tr}\nolimits}

\newcommand{\zhr}{{\zh r}}

\newcommand{\zhn}{{\zh n}}
\newcommand{\zhp}{{\zh p}}

\newcommand{\zhsigma}{{\zh\sigma}}
\newcommand{\zhzeta}{{\zh\zeta}}

\newcommand{\unin}{\hat{\zh n}}

\newcommand{\hXi}{{\Delta\hat{V}}}

\newcommand{\Eq}[1]{Eq.\ (\ref{#1})}

\newcommand{\Fig}[1]{Fig.\ \ref{#1}}
\begin{document}	
	\title{Electron polarization in the resonant inelastic scattering on hydrogen-like ions}
	\author{ D.\ M.\ Vasileva$^{1}$, K.\ N.\ Lyashchenko$^{1,2}$, O.\ Yu.\ Andreev$^{1,2}$}
	\address{$^1$ Petersburg Nuclear Physics Institute named by B.P. Konstantinov of National
		Research Centre "Kurchatov Institute", Gatchina, Leningrad District 188300, Russia}
	\address{$^2$ Department of Physics,
		St.\ Petersburg State University, 7/9 Universitetskaya nab.,
		St. Petersburg, 199034, Russia}
	\ead{summerdacha64@gmail.com}
	%
	\begin{abstract}
We investigate the polarization of the electron beam acquired during the inelastic resonant scattering on hydrogen-like ions initially being in the ground state. The formation and subsequent Auger decay of the intermediate $(3l3l')$ doubly excited states in the resonant channel modify the mechanism of polarization change by enhancing both spin-orbit and exchange interactions. Consequently, in the presence of the resonant channel, the acquired polarization can be clearly observed even for light ions when it is challenging to discern which state of the ion was excited in the process. We also show that the energy dependence of the polarization parameter clearly demonstrates strong interference both between the contributions of specific
autoionizing states in the resonant channel and between the non-resonant and resonant channels.
	\end{abstract}	
    
	\maketitle{}
	\section{Introduction}
    The processes occurring in atom-ion and electron-ion collisions have typically been
    investigated with the unpolarized electron and ion beams. However, such spin-unresolved studies offer only a limited perspective. With the advancement of polarized electron beam production and analysis, a more comprehensive analysis that accounts for the polarization states of both the electron and the ion during the scattering process is now available \cite{Vockert2019,PSTP2024}. Such an analysis can facilitate a deeper understanding of the underlying dynamics of the collision and potentially provide new insights into the investigation of atomic structure. One of the issues of particular interest lies in identifying circumstances under which nuclear polarization becomes a factor in atomic reaction, thus facilitating access to another tool that can offer a glimpse into the nucleus.
    
    Polarized electron beams are widely used in various fields of science and
    technology, such as particle physics, atomic physics and materials science \cite{Lin2023}. The process of
    changing the polarization of an electron beam as it passes through various media and
    interacts with other particles is a fundamental phenomenon of quantum mechanics.
    
    The majority of research into the behavior of electron polarization has been focused on the following areas: the scattering of electrons by atoms and molecules, interactions with the crystal lattice of a solid, and movement of electrons in a magnetic field \cite{Kanik2023}. The electron polarization change in collisions
    with ions, in particular with highly charged ions, has been studied to a lesser
    extent. However, a detailed knowledge of spin-dependent effects is crucial for the comprehensive understanding of the processes occurring in ionized media and the interaction of charged particles with matter.
    
    The polarization changes that occur during a collision are mainly a result of two
    factors: the spin-orbit interaction, which takes place whenever an electron is moving
    in the electric field of the ion, and the spin exchange between the electrons that are
    participating in the collision \cite{kessler,burke2011b}. The influence of
    nuclear spin is minimal and is thus excluded from this analysis \cite{Hanne83}.
    
    In this work, we are concerned with the inelastic scattering of electrons, whereby an
    incident electron transfers a portion of its energy to an ion, resulting in the
    formation of an excited state. As a mechanism for the production of ions in excited
    states, this process (also known as electron impact excitation) is of particular
    significance for the study of plasmas. In particular, the light emitted in the radiative
    decay of excited states of highly charged ions represents a valuable source of
    information about a range of astrophysical objects \cite{astroPerseus}.
    
    The polarization of the emitted light from the excited $2p$ states of hydrogen-like and
    helium-like ions formed in the process of inelastic scattering has been the subject of
    considerable study \cite{Nakamura01,Robbins06,Bostock09,Chen15,Wu22,Wu23}. However, to the best of our knowledge, only the non-resonant
    channel of inelastic scattering has been considered. Our aim is to examine the
    resonant channel, in which an intermediate doubly excited state of the
    corresponding helium-like ion is formed and subsequently undergoes Auger decay. The formation of these intermediate states results in a prolonged collision time. Consequently, the incident electron spends a greater amount of time in the strong electromagnetic field near the nucleus and also interacts more strongly with the bound electron. This results in the enhancement of both spin-orbit and exchange interactions in the resonant channel, which, in turn, gives rise to substantial alterations in the polarization behavior at energies where the collision system energy is relatively close to the energy of one or more doubly excited states of the helium-like ion.
    In particular, in our
    previous work \cite{Vasileva2021}, we explored this effect in resonant
    elastic collision with H-like B${}^{4+}$, Ca${}^{19+}$ and Kr${}^{35+}$ ions.
    
    In this study, we examine the process in which the $(3l3l')$ autoionizing states (where $l$
    is the orbital moment) participate in the scattering, resulting in the excitation of the
    H-like ion to the $2s$, $2p_{1/2}$, and $2p_{3/2}$ states. For the $2s$ final state, the polarization change in the non-resonant channel is almost entirely determined by the spin-orbit interaction. As a purely relativistic effect, it is anticipated to be rather small even for medium $Z$ ions. Conversely, for
    the $2p$ states, the non-resonant component is enhanced due to fine structure
    splitting \cite{Hanne83}. In the energy range under consideration, the energy transfer in the
    scattering process exceeds 95\% of the incident electron kinetic energy, indicating a strongly
    inelastic regime where non-resonant polarization change for the $2p$ final states is
    the largest.
    
    To describe the polarization properties, we calculate spin-dependent amplitudes
    employing an \textit{ab initio} QED treatment for inelastic scattering along with the line-profile approach (LPA) for the description of intermediate doubly-excited states \cite{andreev08pr}. This
    method directly accounts for both relativistic and QED effects, including the Breit
    interaction. The calculations are made for scattering on H-like F${}^{8+}$, Ca${}^{19+}$ and
    Kr${}^{35+}$ ions.
    
    The relativistic units are used throughout the paper.	
	\section{Theory}
	It is established that the full set of polarization properties can be derived from a complete knowledge of spin-dependent amplitudes \cite{kessler}. To calculate the relevant amplitudes, we employ an \textit{ab initio} method described in detail in \cite{Vasileva2024,res2020}. Below we outline the main ideas of our approach and discuss its application to the present study. 
	
	The inelastic scattering process can be schematically depicted as follows:
	\begin{eqnarray}
		e^-_{\zhp_i}+X^{(Z-1)+}(1s) \longrightarrow e^-_{\zhp_f}+X^{(Z-1)+}(2l)\,. \label{nonreschannel}
	\end{eqnarray}
    In the initial state, the ion has a single electron occupying the ground $1s$ state. The ion is subsequently excited to a higher state by the incident electron. In this study, we consider the final states of the ion with the main quantum number $n=2$, that is, the $2s$, $2p_{1/2}$ and $2p_{3/2}$ states. We denote the incident and scattered electron momenta $\zhp_i$ and $\zhp_f$, respectively.
		
	We start with the wave functions $\Psi^i_{m_i,\mu_i^{(\zhzeta_i)}}(\zhr_1,\zhr_2)$ and $\Psi^f_{m_f,\mu_f^{(\zhzeta_f)}}(\zhr_1,\zhr_2)$ describing the initial and final states of the two-electron system in the zeroth order of the perturbation theory:
	\begin{eqnarray}
		\Psi^i_{m_i,\mu_i}(\zhr_1,\zhr_2)&=& \frac{1}{\sqrt{2}}\det{\lbrace\psi^{(+)}_{\zhp_i\mu_i}(\zhr_1),\psi^{(1s)}_{m_i}(\zhr_2)\rbrace} \label{psii},\\
		\Psi^f_{m_f,\mu_f}(\zhr_1,\zhr_2)&=& \frac{1}{\sqrt{2}}\det{\lbrace\psi^{(-)}_{\zhp_f\mu_f}(\zhr_1),\psi^{(2l)}_{m_f}}(\zhr_2)\rbrace \label{psif},
	\end{eqnarray}
	where $\psi^{(1s)}_{m_i}(\zhr)$ and $\psi^{(2l)}_{m_f}(\zhr)$ are the wave functions describing the bound electron in the hydrogen-like ion with the total angular momentum projections on $z$-axis $m_i$ and $m_f$; $\psi^{(\pm)}_{\zhp\mu^{(\zhzeta)}}(\zhr)$ is the in- $(+)$ or out-going $(-)$ wave function of an electron in the electric field of the atomic nucleus with the asymptotic
	momentum $\zhp$ \cite{akhiezer65b}:      
	\begin{eqnarray}
		\psi^{(\pm)}_{\zhp\mu}(\zhr)=\frac{(2\pi)^{3/2}}{\sqrt{p\varepsilon}}\sum_{jlm}\Omega^{+}_{jlm}(\hat{\zhp})v_{\mu}^{(\zhzeta)}e^{\pm i\phi_{jl}} i^l \psi_{\varepsilon jlm}(\zhr)\,, \label{psi_asympt}
	\end{eqnarray}
	where $\psi_{\varepsilon jlm}(\zhr)$ is the continuum solution of the Dirac equation with certain parity, total angular momentum and its projection, $\phi_{jl}$ is the phase shift \cite{akhiezer65b}, $\Omega_{jlm}(\hat{\zhp})$ are the spinor spherical harmonics \cite{varshalovich} 
	and the
	spinor $v_{\mu}^{(\zhzeta)}$  for a given direction defined by the unit vector $\hat{\zhzeta}$
	is determined by the following equation:
	\begin{equation}
		\frac{1}{2} (\hat{\zhzeta}\zhsigma)v_{\mu}^{(\zhzeta)}=\mu v_{\mu}^{(\zhzeta)}\,,
	\end{equation}
    where $\zhsigma$ is the vector constructed with the Pauli matrices $\sigma_x$, $\sigma_y$ and $\sigma_z$.
    The incident and scattered electron wave functions with varying directions of $\hat{\zhzeta}$ (\Eq{psi_asympt}) are employed within our method to yield all pertinent spin-dependent amplitudes.
	
	Within our approach, the amplitude of the inelastic scattering process can be written as follows \cite{Vasileva2024}: 
	\begin{eqnarray}
		U_{if}(m_i,\mu_i,\hat{\zhzeta_i},m_f,\mu_f,\hat{\zhzeta_f}) 
		=
		\langle\Psi^f_{m_f,\mu_f}(\zhr_1,\zhr_2)|\hXi|\Phi^i_{m_i,\mu_i}(\zhr_1,\zhr_2)\rangle
		\,, \label{ampl}
	\end{eqnarray}
    where $\Psi^f_{m_f,\mu_f}(\zhr_1,\zhr_2)$ is the zeroth order final state wave function (\Eq{psif}), $\hXi$ is the operator describing scattering process. In this work, only the one-photon exchange corrections were included in $\hXi$. Wave function $\Phi^i_{m_i,\mu_i}(\zhr_1,\zhr_2)$ can be presented as the sum of two terms:
    \begin{eqnarray}
    	\Phi^i_{m_i,\mu_i}(\zhr_1,\zhr_2) \nonumber
    	&=&
    	C_0\Psi^i_{m_i,\mu_i}(\zhr_1,\zhr_2)\\
    	&+&\sum_{n_1,n_2 \in \mathbf{bound}}
    	C_{n_1 n_2}\Psi^{(0)}_{n_1 n_2}(\zhr_1,\zhr_2)+\cdots
    	\,.	\label{LPA_exp}
    \end{eqnarray}
	The coefficient $C_0$ in the first term is nearly equal to unity. Accordingly, the first term is essentially identical to the initial state zeroth order wave function (\Eq{psii}) and corresponds to the non-resonant channel. The second term $\sum
	C_{n_1 n_2}\Psi^{(0)}_{n_1 n_2}(\zhr_1,\zhr_2)$ is orthogonal to the first and contains the admixture of the bound states of the two-electron
	ion describing the formation of the intermediate autoionizing states in the resonant
	channel:
	\begin{eqnarray}
		e^-_{\zhp_i}+X^{(Z-1)+}(1s) &\longrightarrow& X^{(Z-2)+}(n_1 n_2) \nonumber
		\\
		&&\nonumber
		\hspace{30pt}\downarrow
		\\
		&&
		e^-_{\zhp_f}+X^{(Z-1)+}(2l)\,. \label{reschannel}
	\end{eqnarray}
	
	The coefficients $C_{n_1 n_2}$ in \Eq{LPA_exp} are derived using the LPA for quasidegenerate states described in \cite{andreev08pr}. 
	The method accounts for the interaction with the atomic nucleus and relativistic effects, including spin-orbit interaction, in all orders of perturbation theory. Within the LPA, the dominant part of the interelectron interaction within a selected set of two-electron states is summed in all orders of perturbation theory. In this work, the interelectron interaction is taken into account in all orders for electrons with main quantum number $n\leq5$, while the remaining interelectron interaction is treated within standard QED perturbation theory. Additionally, our calculation incorporates the electron self-energy and vacuum polarization corrections, which become important for heavier ions. In particular, the imaginary parts of the radiative corrections contribute substantially to the total widths of the two-electron states. We give the relative contributions of Auger and radiative decay to the total widths of the $3l3l'$ autoionizing states for Ca$^{18+}$ and Kr$^{34+}$ in our work on the resonant inelastic scattering differential cross sections \cite{Vasileva2024}.
	
	While more sophisticated methods might be preferable for dealing with interelectron interaction in very light ions, our approach incorporates QED and relativistic effects in a natural manner, making it well suited for medium and high $Z$ ions. Nevertheless, our method performs efficiently even in relatively light systems. A comparison of our results for F$^{8+}$ \cite{Vasileva2024} with both experimental data and R-matrix calculations \cite{Zouros99} shows strong agreement.
	
	The double differential cross section 
	for the inelastic electron scattering reads
	\begin{eqnarray}
		\frac{d^2\sigma_{if}}{d\varepsilon_f d\Omega_f} (\varepsilon_f, \theta_f)
		=
		2\pi |U_{if}|^2 \delta(\varepsilon_f + \varepsilon_{2l} - \varepsilon_i - \varepsilon_{1s})
		\times
		\frac{\varepsilon_i}{p_i}\frac{p_f \varepsilon_f}{(2\pi)^3}
		\,,
	\end{eqnarray}
	where $\varepsilon_i$ and $\varepsilon_f$ are the energies of the incident and scattered electron, respectively. For brevity, the notations $\sigma_{if}$ and $U_{if}$ are used here, although it is assumed that both are spin-dependent as in \Eq{ampl}.
	
	    We are interested in the polarization initially unpolarized electron beam acquires in the inelastic scattering.
	The polarization of the electron beam can be described by the density matrix $\hat{\rho}$ given by the polarization vector $\zhzeta^e$ \cite{akhiezer65b}: 
	    \begin{eqnarray}
	    	\hat{\rho}=\frac{1}{2} (1+\zhzeta^e \zhsigma)\,,
	    \end{eqnarray}
	and $|\zhzeta^e|\leq 1$ gives the polarization degree of the electron beam.
	
	    If the quantization axis $\hat{\zhzeta}$ in \Eq{ampl} is fixed ($\hat{\zhzeta_i}=\hat{\zhzeta_f}=\hat{\zhzeta}$), the amplitudes $U_{if}(m_i,\mu_i,\hat{\zhzeta},m_f,\mu_f,\hat{\zhzeta})$ calculated for all possible polarizations of the incident and scattered continuum electrons ($\mu_i$ and $\mu_f$) and for all possible total angular projections of the initial and final bound state of the ion ($m_i$ and $m_f$) form the scattering matrix $M$ which completely defines the polarization properties of the scattering
	\cite{Tolhoek1954,Burke1974}.
	If the initial state of the system is described by $\hat{\rho}_i$, then the density matrix $\hat{\rho}_f$ of the system after scattering is given by
	    \begin{eqnarray}
	    	\hat{\rho}_f=M\hat{\rho_i}M^+\,. \label{Mprop}
	    \end{eqnarray}
	    
	    The scattering matrix $M$ is a $4 \times 4$ matrix if the final ion state is $2s$ or $2p_{1/2}$ or a $8 \times 4$ matrix if the final ion state is $2p_{3/2}$. 
	    
	    If initially both the electron beam and the ion are unpolarized, then the polarization vector of the scattered electron beam can be written as
	\cite{Vasileva2021PhysRevA.104.052808}
	    \begin{eqnarray}
	    	\zhzeta^e&=& \frac{\trace(MM^+\zhsigma^{(1)})}{\trace(MM^+)} \,, \label{poldef}
	    \end{eqnarray}
	    where $\zhsigma^{(1)}$ is assumed to be the direct product of $\zhsigma$ corresponding to the continuum electron and the identity matrix of appropriate dimensions ($2 \times 2$ if the final ion state is $2s$ or $2p_{1/2}$ and $4 \times 4$ if the final ion state is $2p_{3/2}$).
	    
	    If an unpolarized electron beam is scattered on an unpolarized ion, the total polarization after scattering integrated over all directions of the scattered electron momentum must also be equal to zero. Due to the geometry of the scattering, any non-zero polarization component in the plane of scattering violates this condition. Therefore, as in the case of the elastic scattering on ions and atoms \cite{johnson1961,Burke1974,Vasileva2021PhysRevA.104.052808}, the polarization acquired by the electron beam in the scattering must be directed along the direction $\zhn$, where
	    \begin{eqnarray}
	    	\zhn= \frac{[\zhp_i \times \zhp_f]}{|[\zhp_i \times \zhp_f]|} \,.
	    \end{eqnarray}
	    Since the direction of the polarization vector is known, it is convenient to describe the polarization of the scattered electron beam by the parameter $P$ also known as Sherman function \cite{sherman56} -- the projection of the polarization vector on the direction $\zhn$ perpendicular to the plane of scattering:    
	    \begin{eqnarray}
	    	P=\zhzeta^e \cdot \zhn \,. \label{Sdef1}
	    \end{eqnarray}
	    The possible values of parameter $P$ lie between $-1$ and $1$, the absolute value $|P|$ gives the polarization degree of the electron beam.
	We can also write the parameter $P$ as the difference between the cross sections with the opposite scattered electron spin projections on $\unin$ divided by the total cross section
	\cite{Vasileva2021PhysRevA.104.052808}
	    \begin{eqnarray}
	    	&&P= \frac{\left( \frac{d\sigma}{d\Omega}\right) _{0 \rightarrow \zhn}-\left( \frac{d\sigma}{d\Omega}\right) _{0 \rightarrow -\zhn}}{\left( \frac{d\sigma}{d\Omega}\right) _{0 \rightarrow \zhn}+\left( \frac{d\sigma}{d\Omega}\right) _{0 \rightarrow -\zhn}} \,, \label{Sdef2}
	    \end{eqnarray}
    where $\left({d\sigma}/{d\Omega}\right) _{0 \rightarrow \hat{\zhzeta}}$ is the cross section of the process in which an initially unpolarized electron beam has a projection of polarization of $1/2$ on the direction $\hat{\zhzeta}$ after scattering.
	
	Both expressions given by \Eq{poldef} and \Eq{Sdef2} can be used for numerical calculations within our approach. While the former is more convenient for the purpose of calculation, the latter offers a more intuitive representation and serves as an effective verification of the results.

\section{Results and discussion}
The subject of this study is the polarization that an initially unpolarized electron beam acquires after scattering on an unpolarized ion. In order to describe the polarization acquired by the electron beam, we examine the polarization parameter $P$, as defined by \Eq{Sdef1} and \Eq{Sdef2}.

The method outlined in the previous section was developed with a particular focus on scattering on medium and heavy ions. In these systems, relativistic and particularly QED effects, as well as the complete incorporation of the Breit interaction, are of particular importance. Accordingly, Kr$^{35+}$ and Ca$^{19+}$ were selected to illustrate our findings. 

\begin{figure}
	\centering
	\includegraphics[width=0.95\textwidth]{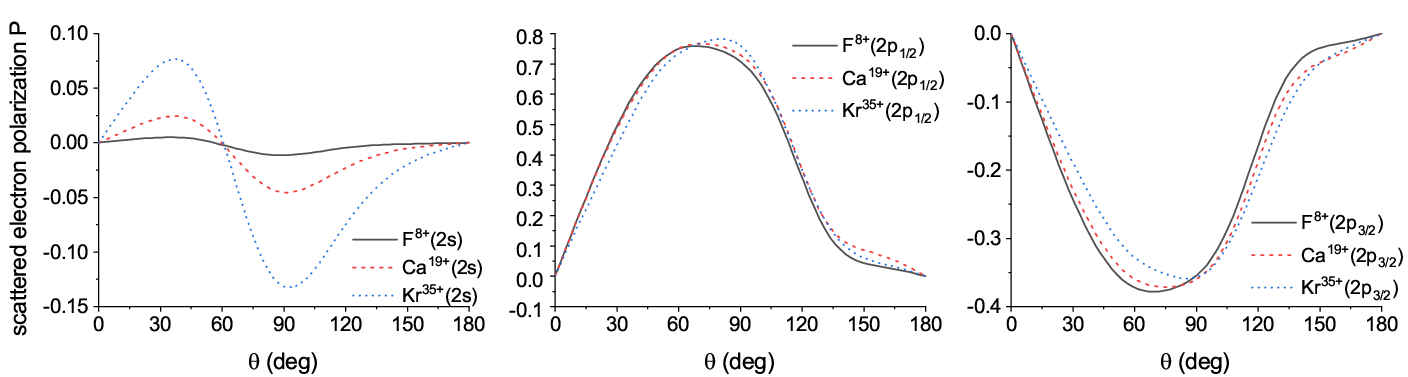}
	\caption{The polarization parameter $P$ for the non-resonant scattering on F$^{8+}$, Ca$^{19+}$ and Kr$^{35+}$ for the final states $2s$, $2p_{1/2}$ and $2p_{3/2}$ as a function of the angle between the incident and scattered electrons. The kinetic energy of the scattered electron is 60 eV for F$^{8+}$, 225 eV for Ca$^{19+}$ and 700 eV for Kr$^{35+}$.}
	\label{fig1}
\end{figure}

The essential characteristic of inelastic scattering is that the contributions of the non-resonant and resonant channels both decrease with increasing atomic number $Z$, allowing resonances to remain highly visible. Still, the experiment can be challenging since the total cross section decreases. For comparison, the total non-resonant cross section for the energies in the range under consideration are 3.74 kBarn for F$^{8+}$, 1.47 kBarn for Ca$^{19+}$ and only 0.13 kBarn for Kr$^{35+}$ (the final electron energies $\varepsilon_f$ are chosen the same as in \Fig{fig1}).

The existing experiments on the inelastic scattering were conducted using light ions \cite{Toth96,Grabbe97,Zavodszky99,Zouros99,Patrick_Richard_1999}. For instance, inelastic scattering in the collision of molecular hydrogen with F$^{8+}$ was studied in \cite{Zouros99}. As previously demonstrated \cite{Vasileva2024}, our approach can be successfully applied to such systems. This section will conclude with a discussion of the scattering on the F$^{8+}$ ion.

As our focus is on the resonant process, the energy spectrum of the incident and scattered electron beam is defined by the following:
\begin{eqnarray}
    \varepsilon_i+\varepsilon_{1s}=\varepsilon_f+\varepsilon_{2l} \approx E^{(i)}_{3l3l'}\,, \label{econs}
\end{eqnarray}
where $E^{(i)}_{3l3l'}$ are the energies of the $(3l3l')$ doubly excited states of the corresponding helium-like ion. This suggests that resonant scattering occurs near the energy threshold of the electron impact excitation, with over 95\% of the incident electron energy used to excite the ion.

The energies $E^{(i)}_{3l3l'}$ of $(3l3l')$ autoionizing states for F$^{7+}$, Ca$^{18+}$ and Kr$^{34+}$ are presented in our previous paper \cite{Vasileva2024}.

In order to establish a point of reference for the involvement of autoionizing states, it is first necessary to consider the non-resonant background. Given that the background remains largely unaltered within the specified energy range, it is sufficient to present the data for a single energy value. For the sake of convenience, the energies for the calculation were chosen to be slightly larger than the energy of the highest lying resonance. \Fig{fig1} illustrates the polarization of the scattered electron beam as a function of angle $\theta$ between incident and scattered electron momenta for three possible final states of the F$^{8+}$, Ca$^{19+}$ and Kr$^{35+}$ ions. It is immediately evident that the dependence of the acquired polarization on the atomic number $Z$ is qualitatively different for the cases of the $2s$ and $2p$ final states.
	
In the non-resonant channel, the polarization behavior can be fairly accurately explained by the sudden approximation \cite{Hanne83}. In this case, the characteristic time of the momentum transfer between the incident electron and ion (i.e. collision time) is several orders of magnitude smaller than that of the spin-orbit interaction. Therefore, these interactions can be considered as acting consecutively and independently. The spin-orbit interaction is a relativistic effect, and for the continuum electron remains rather small even for medium $Z$ ions. In collisions of unpolarized electron beams with unpolarized ions, the exchange interaction by itself  cannot produce any polarization. Thus, for the final $2s$ state, the change in polarization is primarily driven by the spin-orbit interaction, resulting in a quadratic dependence of the acquired polarization on atomic number $Z$. This differs only marginally in qualitative terms from the polarization change observed in Mott scattering \cite{johnson1961}, albeit with a more complex angle dependence. For light ions, the polarization change is essentially negligible.

In contrast, the polarization acquired with the excitation of 2p states is largely independent of $Z$ and rather substantial, due to an effect that occurs when energy levels with fine structure splitting are excited \cite{kessler}. When exchange interaction happens during collision, $2p_{1/2}$ and $2p_{3/2}$ states are not resolved. After that, ion relaxation happens, which follows statistical rules so that the following equalities hold true with a high degree of accuracy \cite{Hanne83}:
\begin{eqnarray}
 P_{2p_{1/2}}=-2P_{2p_{3/2}} \,,\\ \label{zeropol}
 \left( \frac{d\sigma}{d\Omega d\varepsilon_f}\right)_{2p_{1/2}}= \frac{1}{2}\left( \frac{d\sigma}{d\Omega d\varepsilon_f}\right)_{2p_{3/2}} \,.
\end{eqnarray}
While the total polarization change caused by the exchange interaction remains zero, the separate contributions of $2p_{1/2}$ and $2p_{3/2}$ channels are quite substantial. If the processes with different
final states of the ion can be distinguished (for example,
if the fine structure difference is sufficiently large), the
polarization change can be observed.

\begin{figure}
	\centering
	\includegraphics[width=0.95\linewidth]{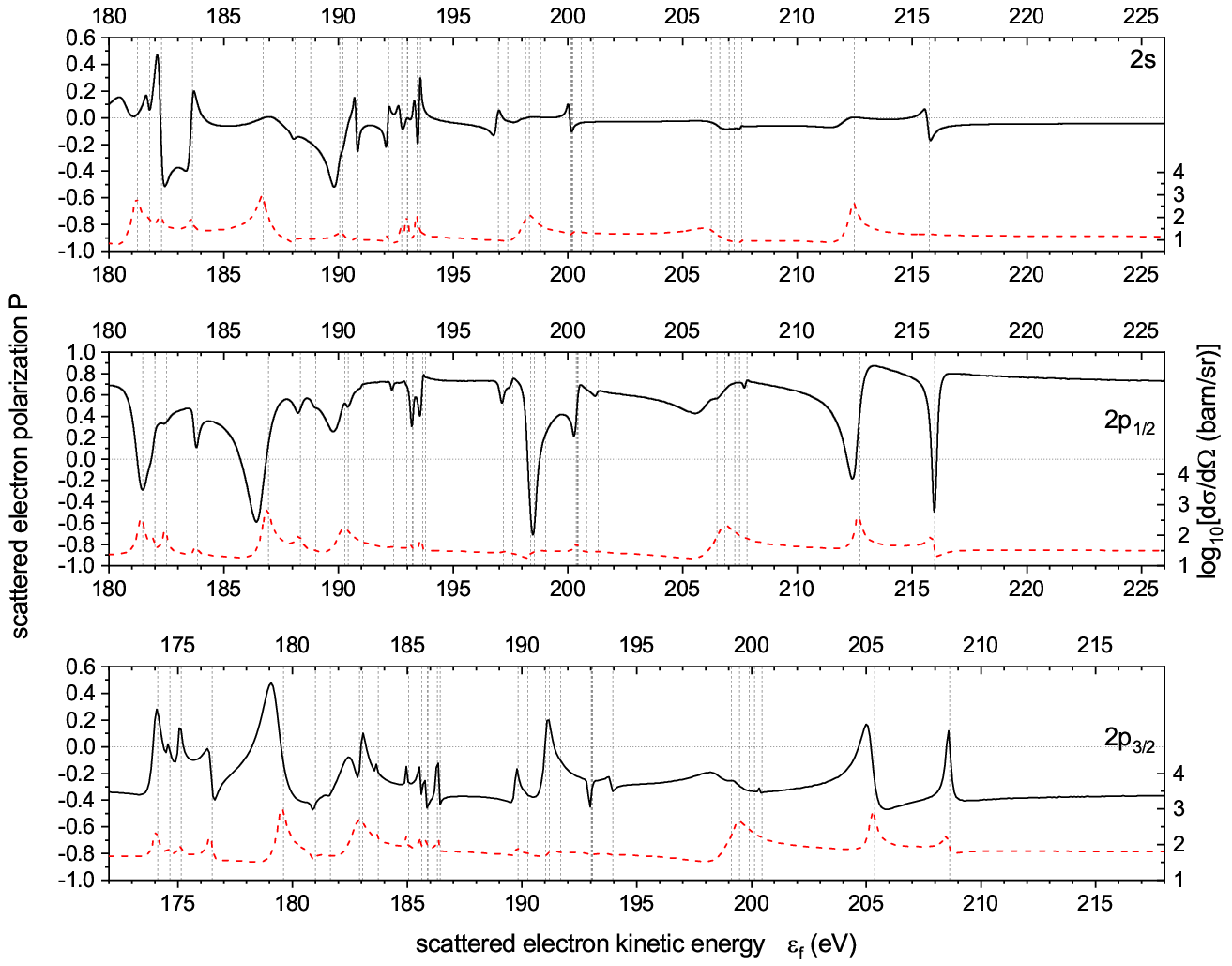}
	\caption{The polarization parameter $P$ of the scattered electron after excitation of Ca$^{19+}$ ion into the $2s$, $2p_{1/2}$ and $2p_{3/2}$ states, respectively, as a function of the scattered electron kinetic energy $\varepsilon_f$ for the polar angle $\theta=90\degree$. The red dashed line shows the decimal logarithm of the corresponding differential cross section. Positions of the resonances are indicated by the vertical gray dashed lines.}
	\label{Ca_90}
\end{figure}

If the energy of the incident electron is such that the energy of the initial state is close to that of an autoionizing state (\Eq{econs}), the resonant channel (\Eq{reschannel}) becomes active.  In this channel, the scattering occurs via the formation and subsequent decay of an intermediate bound state of the corresponding two-electron ion, resulting in a significant alteration of the electron polarization change process due to the facilitation of spin exchange between two electrons and the enhancement of spin-orbit interaction through the trapping of the incident electron in a stronger field for a prolonged period. Both of these effects were previously discussed in \cite{Vasileva2021PhysRevA.104.052808} in the context of resonant elastic scattering. Furthermore, due to a longer collision time, spin-orbit and exchange interaction can no longer be considered as acting independently. Given that our approach inherently includes both spin-orbit and exchange interaction, it is not possible to discern their individual contributions within the scope of this study.

A notable feature of the inelastic scattering is that the resonant channel involves the autoionizing states that are in close proximity to each other in terms of energy. This leads to the simultaneous participation of multiple autoionizing states in the scattering process, resulting in substantial interference between resonances. Additionally, the interference between non-resonant and resonant channels also is integral to the process.

In \Fig{Ca_90} we present the polarization parameter $P$ in the electron scattering on Ca$^{19+}$ ion as a function of the scattered electron kinetic energy for three different excitations of the ion.
The angle $\theta$ between the incident and scattered electron momenta was selected to be 90\degree. The complex resonant structure with seemingly shifted peaks is clearly discernible. In contrast with the differential cross section (for which the decimal logarithm is presented by red dashed lines on the graph), the resonances have Fano-like shapes. This, together with the interference between resonances, leads to the noticeable mismatch between the positions of the resonances (gray dashed vertical lines on the graph) and maxima and minima in the polarization parameter. Accordingly, the peaks of the differential cross section and polarization parameter do not always align. This phenomenon allows for the clear observation of the effects of interference in experiment.

Furthermore, the graph clearly demonstrates that the resonance structure of the polarization parameter displays a similar level of detail as a logarithm of the differential cross section. Consequently, it can provide not only new information about spin-dependent characteristics of the inelastic scattering process, but also additional insight into the structure of the corresponding helium-like ion.

\begin{figure}
	\includegraphics[width=0.99\linewidth]{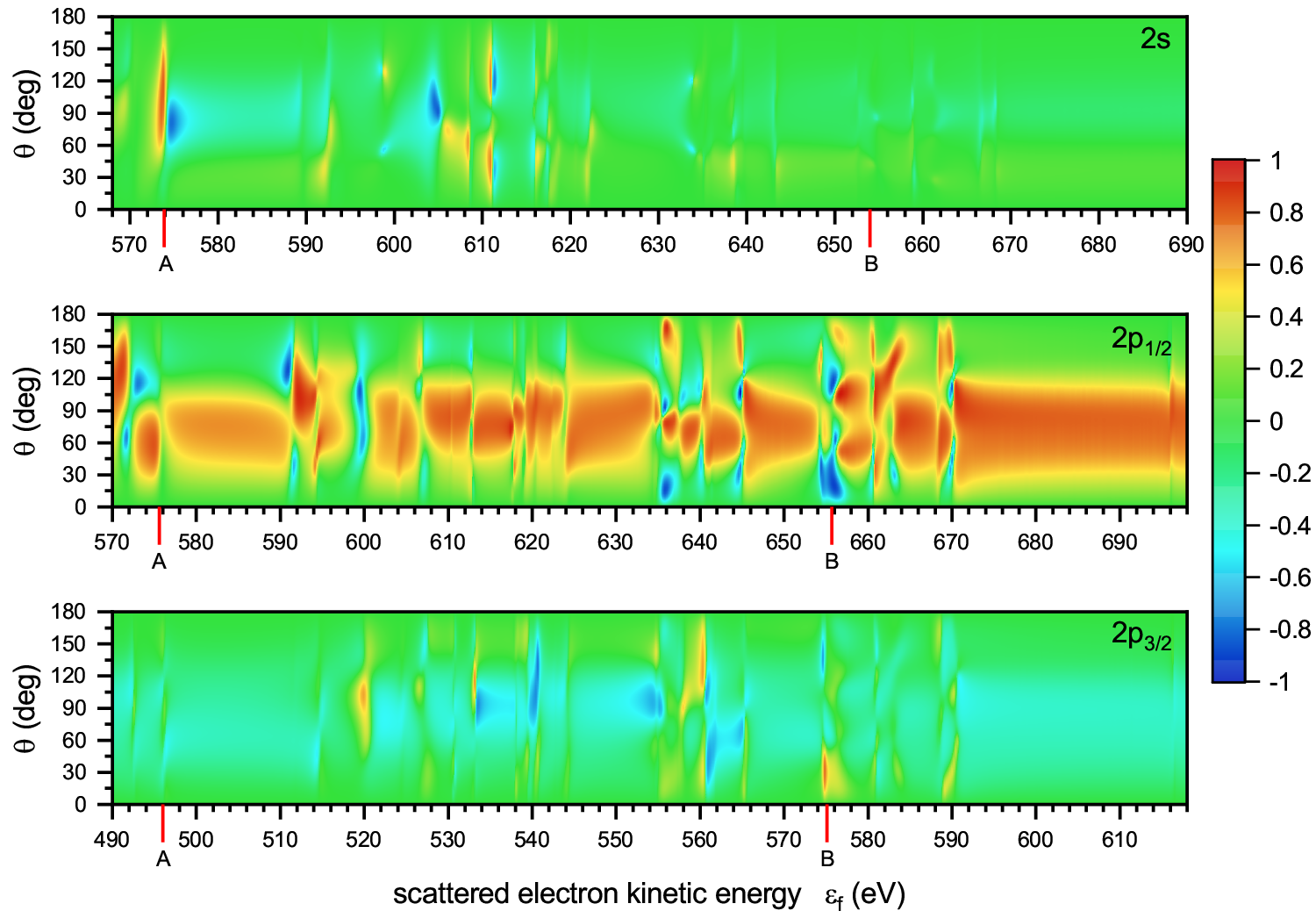}
	\centering
	\caption{The polarization parameter $P$ of the scattered electron after excitation of Kr$^{35+}$ ion into $2s$, $2p_{1/2}$ and $2p_{3/2}$ states, respectively, as a function of the scattered electron kinetic energy $\varepsilon_f$ and polar angle $\theta$. The angular dependence of the polarization parameter P at two selected resonances is shown in more detail on \Fig{res_spec}. The energies corresponding to these resonances are indicated by long red ticks and letters A and B on the energy scale for each of the three scattering channels.}
	\label{Kr_all}
\end{figure}

The complete picture of the polarization parameter is presented in \Fig{Kr_all} for the inelastic scattering on Kr$^{35+}$. The parameter $P$ is demonstrated as a function of both the polar angle and the scattered electron kinetic energy for three potential final ion states. In the bottom two panels ($2p$ states), the non-resonant background is prominent, while the resonant channel in most cases results in a reduction in the absolute value of the polarization parameter $P$. For the $2s$ state, the polarization change is predominantly influenced by the resonant channel, despite the fact that at this $Z$ the background is no longer insignificant. The interference between the background and the resonances is clearly visible. The parameter $P$ exhibits a non-trivial angle dependence at the resonances, which is caused by the interplay between the non-resonant background and multiple resonant contributions. 

\begin{figure}
	\centering
	\includegraphics[width=0.9\linewidth]{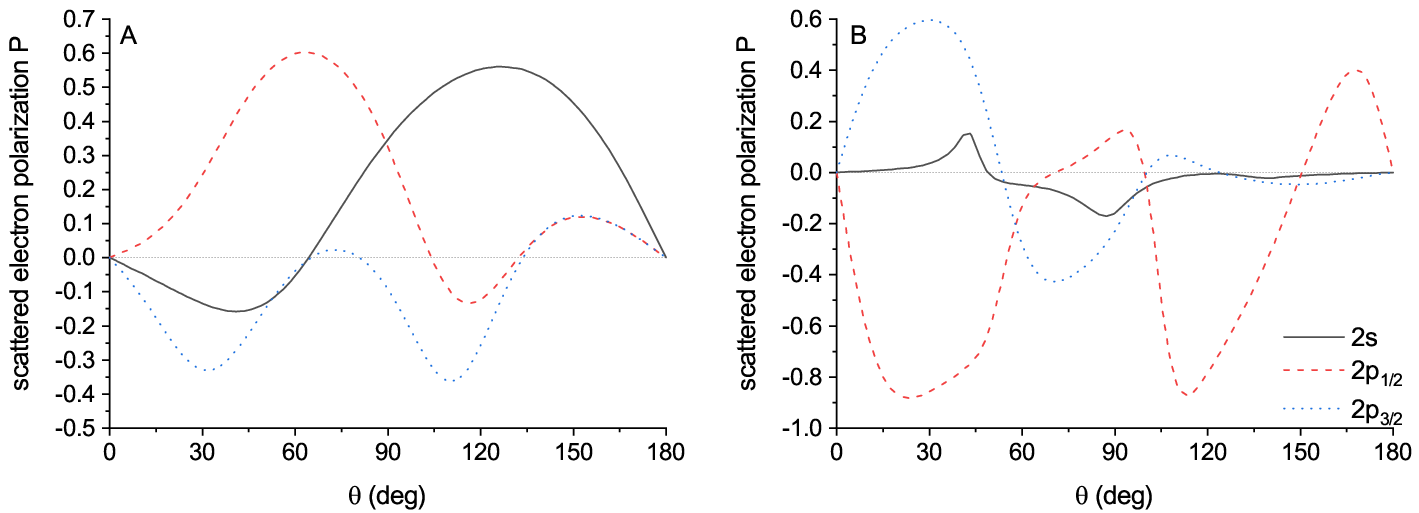}
	\caption{The polarization parameter $P$ of the scattered electron after excitation of Kr$^{35+}$ ion into $2s$, $2p_{1/2}$ and $2p_{3/2}$ states, respectively, as a function of the polar angle $\theta$. In both panels, the energies for three possible channels are chosen so that the same set of autoionizing states is involved in the scattering. The corresponding scattered electron kinetic energies are marked in red in \Fig{Kr_all}}
	\label{res_spec}
\end{figure}

\begin{figure}
	\centering
	\includegraphics[width=0.45\linewidth]{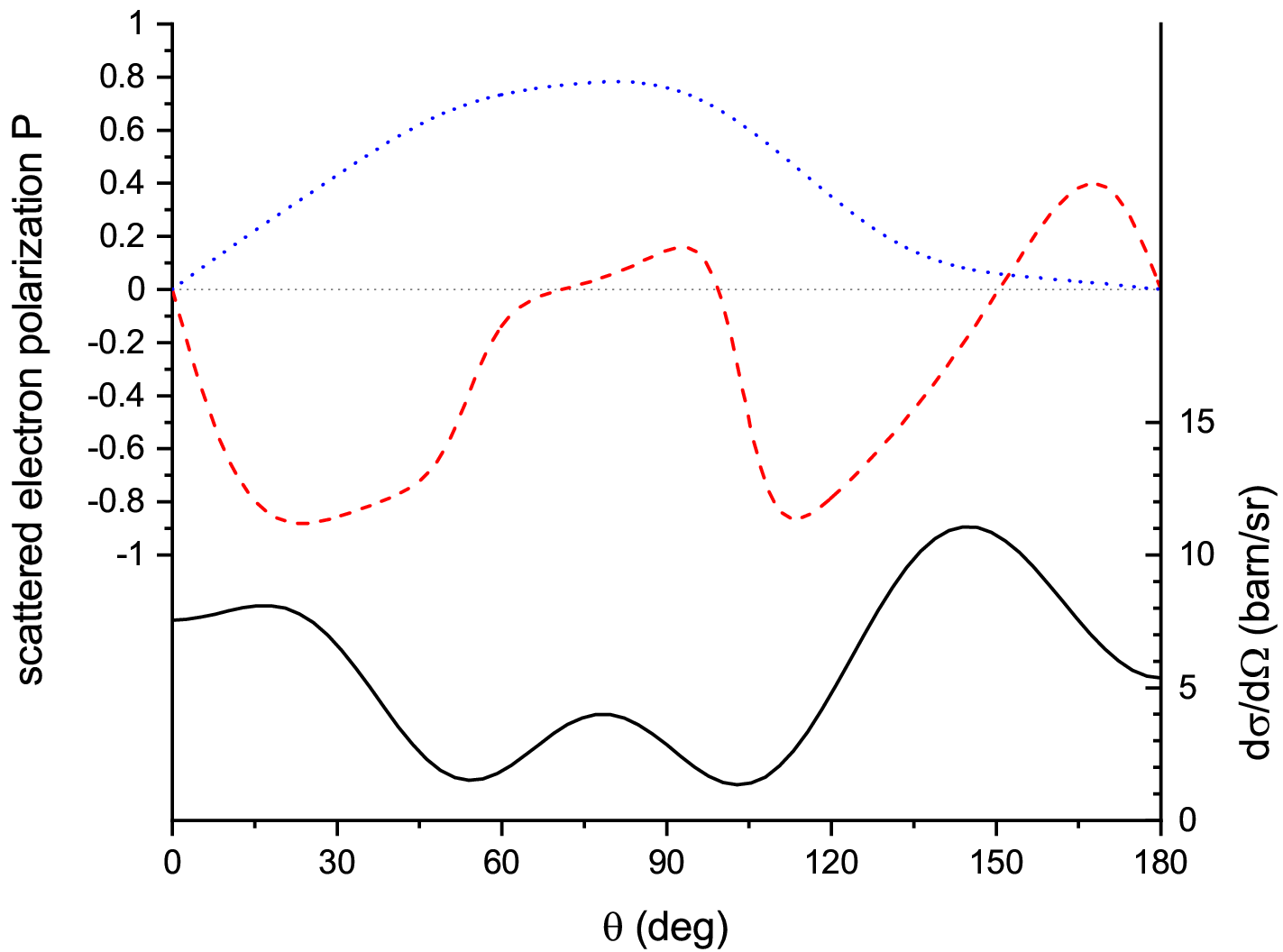}
	\caption{The polarization parameter $P$ of the scattered electron after excitation of Kr$^{35+}$ ion into $2p_{1/2}$ state (red dashed line), the background polarization (the blue dotted line) and the differential cross section (black solid line) as a function of the polar angle $\theta$ for the same energy as in panel B of \Fig{res_spec}}
	\label{res_sc}
\end{figure}

In panels A and B of \Fig{res_spec} the angular dependence of the polarization parameter is shown in more detail for energies A and B marked in red on \Fig{Kr_all}. The energies are chosen  so that they correspond to the same set of intermediate autoionizing states for all possible channels (with excitation of $2s$, $2p_{1/2}$ and $2p_{3/2}$ states). Due to the symmetry of the scattering, the polarization change in forward and backward scattering must be zero. Consequently, the most substantial effect is anticipated for electrons scattered laterally, where the cross section is expected to be minimal. However, at resonance energies this expectation does not always hold, as demonstrated in \Fig{res_sc}. This figure presents the polarization parameter for the $2p_{1/2}$ channel, along with the differential cross section and background polarization. The polarization parameter near the resonance (red dashed line) exhibits a substantial deviation from the non-resonant behavior (blue dotted line), even at relatively small angles. Moreover, in the resonant channel significant polarization change is not limited to the regions where differential cross section (black solid line) is small.

\begin{figure}
	\centering
	\includegraphics[width=0.65\linewidth]{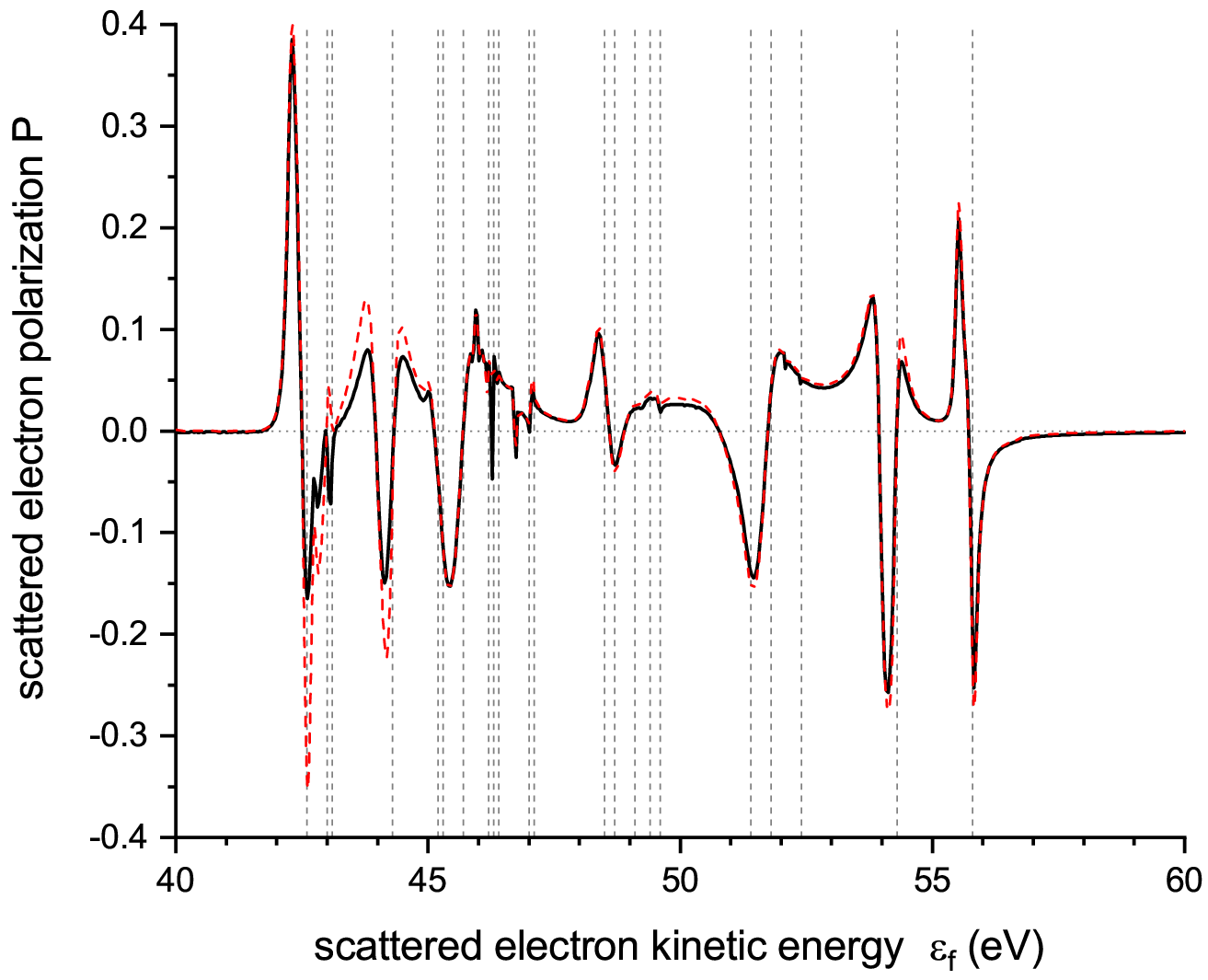}
	\caption{The polarization parameter $P$ (\Eq{P_f}) for the inelastic scattering on F$^{8+}$ at polar angle $90\degree$. The red dashed line shows the contribution of the scattering to $2p$ states. Positions of the resonances are indicated by the vertical gray dashed lines.}
	\label{figF}
\end{figure} 

While the electrons scattered with the simultaneous excitation of ions into $2p_{1/2}$ and $2p_{3/2}$ states can be distinguished by their energies for scattering on Ca$^{19+}$ and to a larger extent Kr$^{35+}$, this is not the case for light ions due to the fine structure splitting being too small. Nevertheless, the resonant channel permits observation of the electron polarization change even in this case as it does not obey \Eq{zeropol}. \Fig{figF} demonstrates the total polarization change
\begin{eqnarray}
    	P_t= \frac{P_{2s} \frac{\phantom{1/} d^2\sigma_{2s_{\phantom{1/2}}}}{d\Omega d\varepsilon_f}  +P_{2p_{1/2}} \frac{d^2\sigma_{2p_{1/2}}}{d\Omega d\varepsilon_f} +P_{2p_{3/2}} \frac{d^2\sigma_{2p_{3/2}}}{d\Omega d\varepsilon_f} }{ \frac{\phantom{1/} d^2\sigma_{2s_{\phantom{1/2}}}}{d\Omega d\varepsilon_f} + \frac{d^2\sigma_{2p_{1/2}}}{d\Omega d\varepsilon_f} + \frac{d^2\sigma_{2p_{3/2}}}{d\Omega d\varepsilon_f} }\,. \label{P_f}
\end{eqnarray}
in the scattering on F$^{8+}$. The red dashed line represents the contribution of $2p$ states only. While the non-resonant background is canceled out, the resonance structure remains clearly visible with peaks reaching a polarization of 40\%. The addition of the $2s$ state results in a slight decrease in the resonance structure. Nevertheless, the acquired polarization remains sufficiently large for observation.


\begin{figure}
	\centering
	\includegraphics[width=0.65\linewidth]{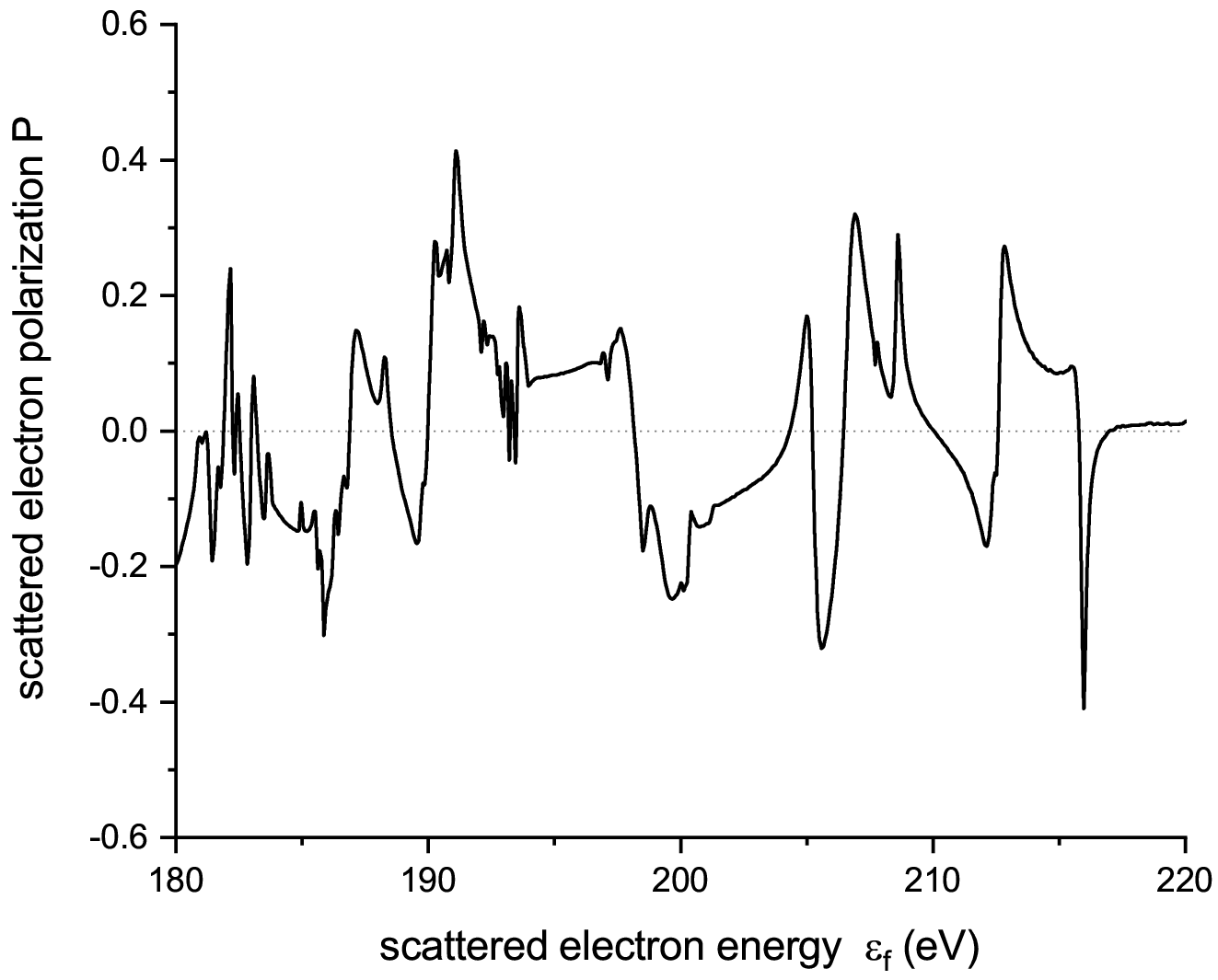}
	\caption{The polarization parameter $P$ (\Eq{P_f}) for the inelastic scattering on Ca$^{19+}$ at polar angle $90\degree$ as a function of scattered electron energy. It is assumed that the energy distribution of the incident electron beam is such that the rate for the scattering channels ($2s$, $2p_{1/2}$ and $2p_{3/2}$) is determined solely by the corresponding differential cross section.}
	\label{figCa_er}
\end{figure}

For heavier ions, in the case where only the energy of the incident electron $\varepsilon_i$ is fixed, a similar picture emerges. However, the situation changes if the scattered electron energy $\varepsilon_f$ is measured. For high $Z$ ions, the energy difference between $2p_{1/2}$ and $2p_{3/2}$ states becomes so large ($81.3$ eV for Kr${}^{35+}$) that the scattered electron energy ranges where resonant channel is effective have only a small overlap. Consequently, the resonant structures corresponding to different channels can be observed separately. In this case, the background is almost canceled out but the resonances stay intact. In the overlap energy region, for all three possible channels to be involved, an initial electron beam with a suitable energy distribution is needed. In principle, the relative rates of three channels depend on the exact shape of the incident electron beam, affecting the resulting shape of the polarization parameter. In order to gain insight into the characteristic polarization of the scattered electron beam, we can consider an ideal case in which the initial energy distribution is such that incident electron beam contributes equally to all three channels. This case is nevertheless quite realistic for atom-ion collision experiments where the initial energy distribution is determined by the Compton profile of the atom which is usually wide compared to the energy range under consideration. Under this assumption, we calculate the total polarization parameter across all channels for scattering on Ca${}^{19+}$  (\Fig{figCa_er}). As previously, a decrease of peak magnitude compared to \Fig{Ca_90} is evident but the polarization change still remains substantial. The overlap of resonances from different channels leads to the broadening of peaks in the resonance structure, which may facilitate experimental observation.	
\section{Conclusions}
We investigated the polarization acquired by the electron beam in the inelastic scattering on F${}^{8+}$, Ca${}^{19+}$ and Kr${}^{35+}$ hydrogen-like ions initially being in the ground state. In the non-resonant energy region of the incident electron, for the final $2s$ state of the ion the polarization acquired by the incident electron beam is determined by the spin-orbit interaction and quadratically grows with atomic number $Z$.  In the case of the $2p$ states, the polarization change is related to the fine-structure splitting of $2p_{1/2}$ and $2p_{3/2}$ states and is largely independent of $Z$.

At resonant energies of incident electrons, the scattering process proceeds through the formation and subsequent Auger decay of autoionization states. The influence of the resonant channel amplifies both spin-orbit and exchange interactions, resulting in the emergence of distinct maxima and minima in the energy dependence of the polarization parameter. The resonances in the polarization parameter typically have Fano-like shapes. The strong interference between resonances results in a misalignment between the peaks in the differential cross section and polarization parameter. The interference between non-resonant and resonant channels is also significant.

In the resonant channel there is no rigid connection between the polarization acquired with the excitation of the $2p_{1/2}$ and $2p_{3/2}$ states. Consequently, the polarization change in the resonant inelastic scattering is clearly observable even in light ions where the small energy difference between states with the main quantum number $n=2$ complicates the detection of the electron scattering with the excitation of a specified state. In turn, highly charged ions provide an opportunity for the independent study of scattering channels with different ion excitations.

\ack
The authors express their gratitude for the hospitality to Prof. Deyang Yu during their visit to the Institute of Modern Physics of the Chinese Academy of Sciences.
The work in part of the calculation of the polarization was supported solely by the Russian Science Foundation under Grant No. 22-12-00043.
The work of K.N.L., O.Y.A. was supported by the Chinese Academy of Sciences (CAS) Presidents International Fellowship Initiative (PIFI) under Grant Nos.
2022VMC0002 and 2025PVA0072, respectively.
\section*{References}    
\providecommand{\newblock}{}

\end{document}